\documentclass[9pt,twocolumn,twoside]{osajnl}

\journal{ol} 

\setboolean{shortarticle}{true}

\title{Detecting momentum weak value: Shack–Hartmann versus a weak measurement wavefront sensor}

\author[1,2]{Yi Zheng}
\author[1,2]{Mu Yang}
\author[1,2]{Zheng-Hao Liu}
\author[1,2]{Jin-Shi Xu}
\author[1,2,*]{Chuan-Feng Li}
\author[1,2]{Guang-Can Guo}

\affil[1]{CAS Key Laboratory of Quantum Information, University of Science and Technology of China, Hefei 230026, China}
\affil[2]{CAS Center for Excellence in Quantum Information and Quantum Physics, University of Science and Technology of China, Hefei 230026, China}

\affil[*]{Corresponding author: cfli@ustc.edu.cn}

\begin{abstract}
The task of wavefront sensing is to measure the phase of the optical field.  Here, we demonstrate that the widely used Shack–Hartmann wavefront sensor detects the weak value of transverse momentum, usually achieved by the method of quantum weak measurement. We extend its input states to partially coherent states and compare it with the weak measurement wavefront sensor, which has a higher spatial resolution but a smaller dynamic range. Since weak values are commonly used in investigating fundamental quantum physics and quantum metrology, our work would find essential applications in these fields.
\end{abstract}

\setboolean{displaycopyright}{true}

\begin{document}

\maketitle

In classical optics, we need a complex amplitude to fully characterize an optical field, which can be regarded as the photon wavefunction in quantum language when we don't need a full quantum description of light \cite{Shih}. The intensity of a stable field is easy to detect, for example, by using a charge-coupled device (CCD), but the phase isn't. People developed various methods to retrieve the phase of light, including the Shack--Hartmann wavefront sensor (SHWS) \cite{PRIMOT200381}. It uses a lens array to focus the incoming oblique light wave within each aperture on a light spot at its focal plane. The position of each spot relative to the center is proportional to the phase gradient $(k_x,k_y)=\nabla \Phi(x,y)=\nabla \arg U(x,y)$ under paraxial approximation, where $U(x,y)$ is the complex amplitude of a monochromatic optical field at a given $z$ \cite{Ares:00}. The phase is the line integral of $(k_x,k_y)$, which can be numerically calculated using methods like zonal and modal estimation \cite{Southwell:80}. In adaptive optics, wavefront sensing is an important tool to correct wavefront distortions.
	
Ordinary quantum measurement forces the system state to collapse, while weak measurement couples the system and the measurement pointer weakly so that the system state doesn't change much. The concept of weak value was proposed by Aharonov \emph{et~al}.\cite{PhysRevLett.60.1351}. It involves an initial state $\left| i\right\rangle$, an observable $A$, and a final state $\left| f\right\rangle$, and the value is $\left\langle A\right\rangle_\textrm{w}=\left\langle f|A|i\right\rangle/\left\langle f|i\right\rangle$, which is generally complex. It is a new way to measure small quantities, determine quantum states, and show quantum paradoxes \cite{RevModPhys.86.307}. Weak or strong \cite{PhysRevLett.105.230401} measurement can be applied to obtain a weak value. The complex amplitude of light has been measured in the context of weak value, including the one-dimensional (1D) method by Lundeen \emph{et~al}.\cite{Lundeen2011} and the 2D scan-free method by Shi \emph{et~al}.\cite{Shi:15}. However, they need either small wave plates or spatial light modulators (SLM) to alter the polarization state of photons located in a small range. The finite size of this range is the main cause of error. Recently, a novel method employing the line integral of the transverse momentum weak value is used for wavefront sensing \cite{Yang2020}. The wavefront sensor avoids the use of Fourier lens and post-selection on the momentum $\boldsymbol{p}=0$ and can be used to measure wavefronts with ultra-high spatial frequency, which we refer to as the weak measurement wavefront sensor (WMWS).
	
In this Letter, we demonstrate that the quantity detected by Shack--Hartmann wavefront sensor also corresponds to the real part of the weak value of transverse momentum. We further derive an expression of the measured quantity of SHWS and WMWS with the input partially coherent light and provide an improvement method of WMWS. Finally, we make a comparison to these two wavefront sensors. Our results would be helpful to investigate fundamental physical problems with robust classical optical devices.

For a pure initial state, the momentum weak value is expressed as
\begin{equation}\label{wvk}
	\left\langle\boldsymbol{k}\right\rangle_\textrm{w}=\frac{\left\langle\boldsymbol{x}\right|\boldsymbol{k}\left|\psi\right\rangle}{\left\langle\boldsymbol{x}|\psi\right\rangle},
\end{equation}
where $\boldsymbol{k}=\boldsymbol{p}/\hbar$. The line integral of $\Re\left\langle\boldsymbol{k}\right\rangle_\textrm{w}$ is the phase of wavefunction
\begin{eqnarray}\label{lineint}
	\Phi(\boldsymbol{x})&=&\arg\psi(\boldsymbol{x})=\Im\ln\psi(\boldsymbol{x})=\Re\left[-\textrm{i}\ln\psi(\boldsymbol{x})\right]  \nonumber    \\
	~&=&\Re\int(-\textrm{i})\nabla\ln\psi(\boldsymbol{x})\cdot\textrm{d}\boldsymbol{s}=\int\Re\left[-\frac{\textrm{i}\nabla\psi(\boldsymbol{x})}{\psi(\boldsymbol{x})}\right]\cdot\textrm{d}\boldsymbol{s} \nonumber    \\
	~&=&\int\Re\frac{\left\langle\boldsymbol{x}\right|\boldsymbol{k}\left|\psi\right\rangle}{\left\langle\boldsymbol{x}|\psi\right\rangle}\cdot\textrm{d}\boldsymbol{s}=\int\Re\left\langle\boldsymbol{k}\right\rangle_\textrm{w}\cdot\textrm{d}\boldsymbol{s}.
\end{eqnarray}
In quantum mechanics, we use density matrix
$\rho=\sum_{i}p_i\left|\psi_i\right\rangle\left\langle\psi_i\right|$ to describe a mixed state
with $p_i$ being the probability of each possible state $\left|\psi_i\right\rangle$. In position space with infinite degrees of freedom (DOF), we define the density matrix function (DMF) in position basis $\rho(\boldsymbol{x}',\boldsymbol{x})=\left\langle\boldsymbol{x}'|\rho|\boldsymbol{x}\right\rangle$. As shown in the literature \cite{Shih,PhysRevLett.105.010401,Stoklasa2014}, $\rho(\boldsymbol{x}',\boldsymbol{x})$ is proportional to the mutual coherence function at the same time
\begin{equation}
	\rho(\boldsymbol{x}',\boldsymbol{x})\propto G(\boldsymbol{x}',\boldsymbol{x};0)=\left\langle U(\boldsymbol{x}',t)U^\ast(\boldsymbol{x},t)\right\rangle,
\end{equation}
and we give an intuitive derivation in Supplement 1. When the initial state is mixed, the weak value is $\left\langle\boldsymbol{x}\right|\boldsymbol{k}\rho\left|\boldsymbol{x}\right\rangle/\left\langle\boldsymbol{x}\right|\rho\left|\boldsymbol{x}\right\rangle$ \cite{PhysRevLett.108.070402}, which becomes \eqref{wvk} when $\rho=\left|\psi\right\rangle\left\langle\psi\right|$. Then the line integral of its real part becomes
\begin{equation}\label{vwrhoint}
	\int\Re\frac{\left\langle\boldsymbol{x}\right|\boldsymbol{k}\rho\left|\boldsymbol{x}\right\rangle}{\left\langle\boldsymbol{x}\right|\rho\left|\boldsymbol{x}\right\rangle}\cdot\textrm{d}\boldsymbol{s}=\int\nabla_1\arg\rho(\boldsymbol{x},\boldsymbol{x})\cdot\textrm{d}\boldsymbol{s},
\end{equation}
whose meaning will be briefly discussed later.

\begin{figure}[h]
\centering\includegraphics[width=6.4cm]{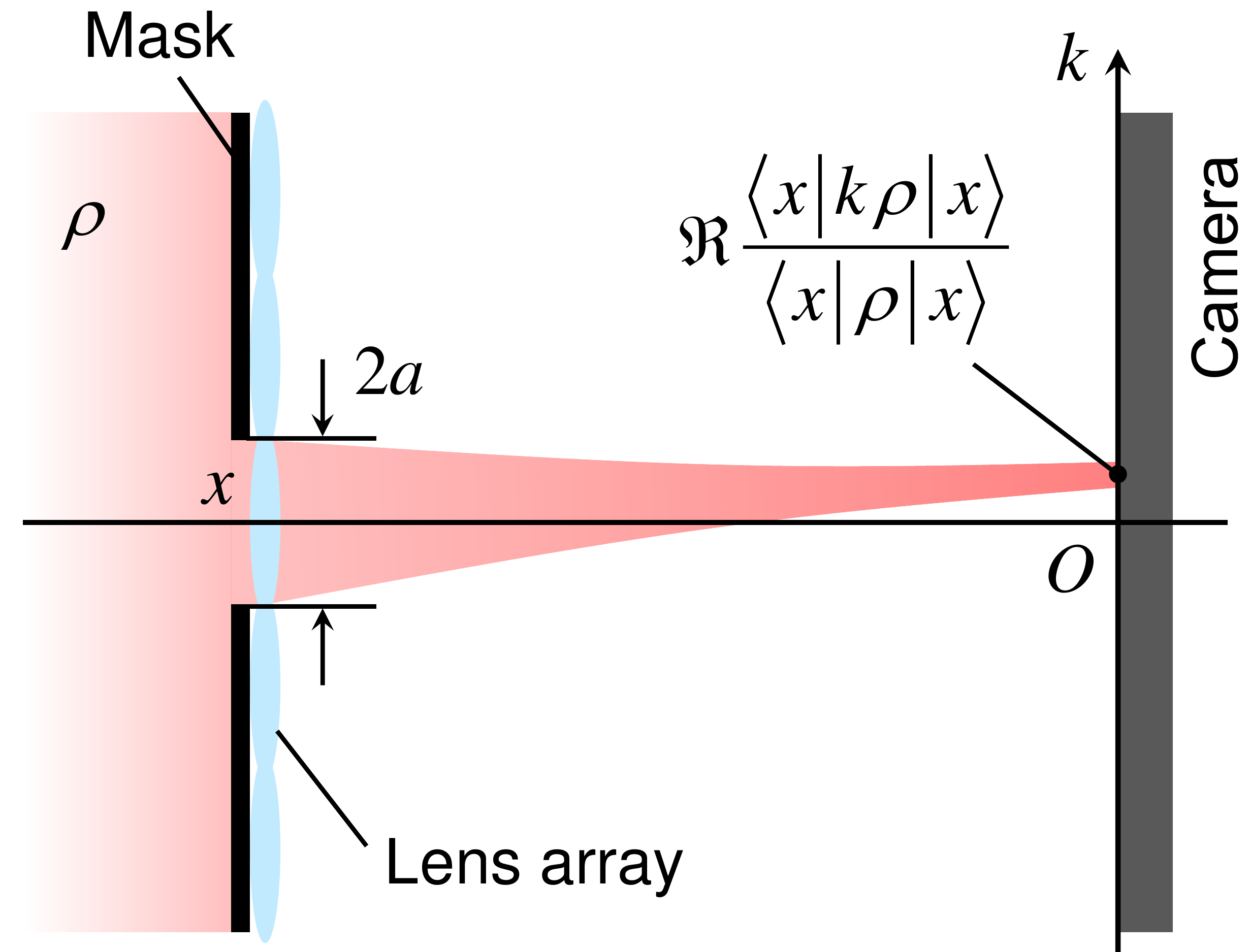}
\caption{Shack--Hartmann wavefront sensor with only one lenslet at position $x$ open. When partially coherent light described by $\rho$ arrives, the center point of light spot on the focal plane corresponds to the real part of momentum weak value.}
\label{p0}
\end{figure}
Detection of partially coherent lights using SHWS and its possible applications have been discussed \cite{PhysRevLett.105.010401,Stoklasa2014}. Now we analyze why SHWS measures momentum weak value. For simplicity, we only consider 1D case and calculate the behavior of light passing the microlens located at $z=0$. Detecting light at other locations is equivalent to displacing the light or DMF first. The input optical field is described by $\rho$, and we let the length of the microlens be $2a$. In a real SHWS, light spots on the focal plane may escape to other zones if the incoming wave is too oblique, causing difficulties when processing data. But here we mask other zones first, as shown in Fig.~\ref{p0} so that there will be only one spot at the focal plane, analogous to short-time Fourier transform \cite{Choi:12}. The quantum operator of masking is a superposition of projection operators, $\Pi=\int_{-a}^{a}\textrm{d}x\left|x\right\rangle\left\langle x\right|$. Applying it to $\rho$, we obtain
\begin{equation}
	\rho_a=\Pi\rho\Pi^\dagger=\int_{-a}^{a}\textrm{d}x_1\int_{-a}^{a}\textrm{d}x_2\left|x_1\right\rangle\left\langle x_1\right|\rho\left|x_2\right\rangle\left\langle x_2\right|,
\end{equation}
which hasn't been normalized under the condition $\operatorname{Tr}\rho_a=1$. SHWS searches the center point of each spot at the focal plane to be the measurement result of this aperture by calculating the intensity-weighted average of positions from the camera image \cite{Ares:00}. The optical field we consider is close to the mask, so at the focal plane an extra phase proportional to $r^2$ is added to the Fourier transform of the original field, but the intensity is not changed. So what SHWS measures is the average momentum of the mixed state $\rho_a$, that is
\begin{equation}
	\frac{\operatorname{Tr}\left(\rho_a k\right)}{\operatorname{Tr}\rho_a}=\frac{\int_{-a}^{a}\textrm{d}x\Im\rho'_1(x,x)}{\int_{-a}^{a}\textrm{d}x\rho(x,x)},
\end{equation}
which is proved in Supplement 1. When $a$ is sufficiently small, we assume $\rho$ and its partial derivatives remain the same in the aperture so that we can replace $\rho(x,x)$ with $\rho(0,0)$, and $\rho'_1(x,x)$ with $\rho'_1(0,0)$, obtaining $\Im\left(\rho'_1(0,0)/\rho(0,0)\right)$. When the aperture is at position $x$, the value is
\begin{equation}\label{vwrho}
	\Im\frac{\rho'_1(x,x)}{\rho(x,x)}=\Re\frac{\left\langle x\right|k\rho\left|x\right\rangle}{\left\langle x\right|\rho\left|x\right\rangle}.
\end{equation}
So when the aperture is small, SHWS detects the real part of momentum weak value.

\begin{figure}[h]
\centering\includegraphics[width=8.2cm]{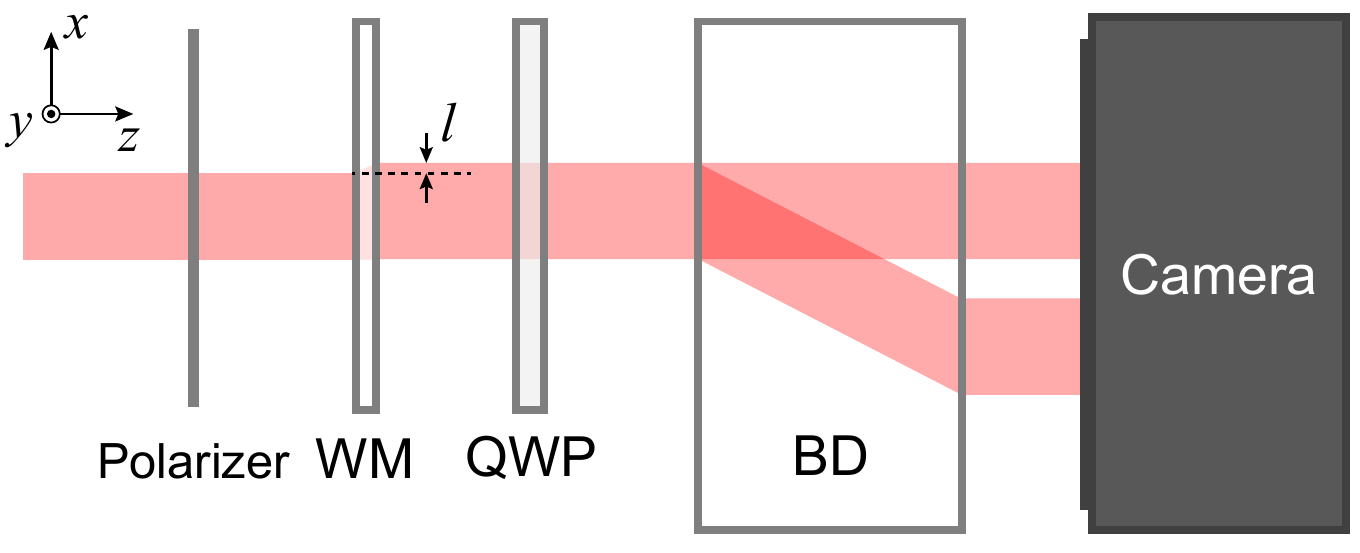}
\caption{Experimental setup of weak measurement wavefront sensor. A beam passes through a polarizer, a calcite crystal (WM), a quarter wave plate (QWP), and a beam displacer (BD). Then the intensity is measured by a camera.}
\label{p1}
\end{figure}
Then we turn to WMWS, which follows the method by Kocsis \emph{et~al}.\cite{Kocsis1170}. As shown in Fig.~\ref{p1}, a polarizer with an optical axis at $45^\circ$ is used to prepare the incoming light beam with diagonal polarization $\left(\left|H\right\rangle+\left|V\right\rangle\right)/\sqrt{2}$. Then it passes a thin calcite crystal called WM which performs weak measurement. The angle between its axis and the propagation direction ($z$-axis) is set to be near $45^\circ$, and the crystal can be rotated around $z$ axis to let its axis lie on either $x$-$z$ or $y$-$z$ plane. When the axis is on $x$-$z$ ($y$-$z$) plane, the incoming light with horizontal (vertical) polarization is displaced by a small distance $l$ toward the positive direction of $x$-axis ($y$-axis). However, this birefringent crystal can't be deemed as a perfect beam displacer. For example, $l$ depends on the oblique angle of the incoming beam, and phase difference of horizontal (\emph{H}) and vertical (\emph{V}) light emerges after passing WM, which needs to be eliminated by tilting WM. See Supplement 1 for more details. Then a quarter wave plate (QWP) and a beam displacer (BD, a thick birefringent crystal) are used to separate beams with left- and right-handed circular polarization. A CCD or CMOS camera measures the intensity of the final beam. Then we can measure the real part of $k_x$ ($k_y$) weak value using this formula
\begin{equation}\label{wvwsf}
	\Re\left\langle k_x\right\rangle_\textrm{w}\approx\frac{1}{l}\frac{I_{x,L}-I_{x,R}}{I_{x,L}+I_{x,R}},\Re\left\langle k_y\right\rangle_\textrm{w}\approx\frac{1}{l}\frac{I_{y,R}-I_{y,L}}{I_{y,L}+I_{y,R}},
\end{equation}
which is an approximate form due to the nature of weak value. An important correction by an arcsine operation \cite{Kocsis1170,Yang2020} will be discussed later.

We provide a theoretical derivation of momentum weak value with the pure input state based on classical optics in Supplement 1. We further describe it using a mixed state $\rho$ when the axis of WM is on $x$-$z$ plane. The $y$-$z$ plane scenario is similar.
	
Considering both spatial and polarization DOFs, the initial density matrix of photon is $\rho_{\textrm{init}}=\frac{1}{2}\left(\left|H\right\rangle+\left|V\right\rangle\right)\left(\left\langle H\right|+\left\langle V\right|\right)\otimes\rho$. The unitary operation of WM is $\exp\left(-\textrm{i}l\left|H\right\rangle\left\langle H\right|\otimes k_x\right)$, which becomes $U_{\textrm{WM}}=\left(1-\textrm{i}l\left|H\right\rangle\left\langle H\right|\otimes k_x\right)$ under first-order approximation when $l$ is small. After passing WM, the state becomes $U_{\textrm{WM}}\rho_{\textrm{init}} U_{\textrm{WM}}^\dagger$. Remember the two circular polarization states $\left|L\right\rangle=\left(\left|H\right\rangle+\textrm{i}\left|V\right\rangle\right)/\sqrt{2}$ and $\left|R\right\rangle=\left(\left|H\right\rangle-\textrm{i}\left|V\right\rangle\right)/\sqrt{2}$. Acting $\left|L\right\rangle\left\langle L\right|$, $\left|R\right\rangle\left\langle R\right|$ separately on it, and omitting polarization DOF, we have
\begin{equation}
	\rho_{x,L/R}=\rho\pm\frac{1}{2}l\left(\rho k_x+k_x\rho\right)+\frac{\textrm{i}}{2}l\left(\rho k_x-k_x\rho\right),
\end{equation}
where the upper symbol represents $L$ and the lower one represents $R$. The intensity at $(x,y)$ is proportional to $\left\langle\boldsymbol{x}|\rho|\boldsymbol{x}\right\rangle$, so
\begin{equation}
	\frac{1}{l}\frac{\left\langle\boldsymbol{x}\right|\left(\rho_{x,L}-\rho_{x,R}\right)\left|\boldsymbol{x}\right\rangle}{\left\langle\boldsymbol{x}\right|\left(\rho_{x,L}+\rho_{x,R}\right)\left|\boldsymbol{x}\right\rangle}\approx\Re\frac{\left\langle\boldsymbol{x}\right|k_x\rho\left|\boldsymbol{x}\right\rangle}{\left\langle\boldsymbol{x}\right|\rho\left|\boldsymbol{x}\right\rangle},
\end{equation}
where the first-order small quantity in the denominator is omitted, proving WMWS measures the same quantity as SHWS in a different way.
	
Till now, we suppose the optical field doesn't diffract, that is, $\rho$ doesn't change during free propagation. In fact, the interaction Hamiltonian commutes with the free-propagation Hamiltonian of photon, so the sensor detects the optical field on the surface of the camera when WM, QWP, and BD are absent \cite{Kocsis1170}.
	
The dynamic range described by the largest detectable oblique angle is a key property of a wavefront sensor. For SHWS, when the focal length is fixed, its dynamic range decreases as each aperture becomes small \cite{Lee:05}. For WMWS, the analysis is more involved. Let $U(x,y)=\exp(\textrm{i}kx)$, and obviously $k_x=k,k_y=0$. But we would obtain $k_x=k\operatorname{sinc}kl$, where $\operatorname{sinc}x=\sin x/x$. The correction method is to take the arcsine of the fraction in \eqref{wvwsf} \cite{Kocsis1170,Yang2020}.
But it's still an approximate result when the amplitude of $U$ is not constant. When $|kl|>\pi/2$, $\sin kl$ is no longer an increasing function, so the dynamic range of WMWS is inversely proportional to $l$. When $l$ decreases, the sensor becomes less perceptive to small $k$ changes, compromising its sensitivity.
	
As WM and QWP are chosen for one wavelength, this wavefront sensor shown in Fig. \ref{p1} is only applicable to lights with a given wavelength. The advent of achromatic wave plate \cite{Beckers:71} provides an idea to make it suitable for a range of wavelengths or non-monochromatic lights. See Supplement 1 for more details.

In this Letter, we've shown Shack--Hartmann wavefront sensor measures the transverse momentum weak value of photons. Other classical optical devices may find their own quantum descriptions in the future, and their measured quantities may be reminiscent of other concepts. This momentum weak value is related to Bohmian velocity \cite{Kocsis1170,LUIS201595,Xiao:17,doi:10.1126/sciadv.aav9547} and probability flux. The measurement result of partially coherent light is presented and the line integral as shown in \eqref{vwrhoint} can be considered as the “equivalent phase as a pure state” as long as $\nabla_1\arg\rho(\boldsymbol{x},\boldsymbol{x})$ is a field with potential. Its applications include measuring the thickness distribution of a transparent specimen illuminated by partially coherent light sources \cite{Gong:17}. As it only concerns $\rho(\boldsymbol{x}',\boldsymbol{x})$ near $\boldsymbol{x}'=\boldsymbol{x}$, it still can't reconstruct the whole DMF. This can be done by performing Fourier transform on Dirac distribution, which can be measured using experimental setup proposed by Bamber and Lundeen (1D case) \cite{PhysRevLett.112.070405}, or scanning $\boldsymbol{p}$ using the setup of Shi \emph{et~al.} (2D case) \cite{Shi:15}.

The weak measurement wavefront sensor is an example of applying quantum ideas to classical optics. We proved that these two sensors measure the same quantity, which means the mature SHWS can also be helpful in measuring Bohmian velocity and trajectory \cite{Kocsis1170}, and WMWS may be applicable for some wavefront sensing tasks, like evaluating Zernike coefficients \cite{Lane:92}. As a supplement to the original letter \cite{Yang2020}, we analyzed its dynamic range and proposed a possible improvement method. Compared with SHWS, it doesn't need lens array, so it's cheaper and the spatial resolution is higher. Its error is mainly from the intensity noise on the camera, rather than the systematic error from weak measurement. To increase its sensitivity, the dynamic range should be as narrow as possible. Its actual performance needs to be experimentally investigated further. If its sensitivity is guaranteed and detection noise is overcome, it may have advantages in detecting minute phase aberrations, for SHWS needs a longer focal length to enhance its sensitivity, thereby increasing the size of Airy spots. Our work provides new perspectives for classical optical detection and quantum optics researches.

\begin{backmatter}
\bmsection{Funding} National Key Research and Development Program of China (Grant No. 2016YFA0302700), National Natural Science Foundation of China (Grants No. 61725504, 61327901, 61490711, 11774335 and 11821404), Key Research Program of Frontier Sciences, Chinese Academy of Sciences (CAS) (Grant No. QYZDY-SSW-SLH003), Science Foundation of the CAS (No. ZDRW-XH-2019-1), Anhui Initiative in Quantum Information Technologies (AHY060300 and AHY020100), Fundamental Research Funds for the Central Universities (Grant No. WK2470000020, WK2470000026 and WK5290000002).

\bmsection{Disclosures} The authors declare no conflicts of interest.

\bmsection{Data availability} No data were generated or analyzed in the presented research.

\bmsection{Supplemental document}
See Supplement 1 for supporting content. 

\end{backmatter}


\bibliography{sample}

\bibliographyfullrefs{sample}

\ifthenelse{\equal{\journalref}{aop}}{%
\section*{Author Biographies}
\begingroup
\setlength\intextsep{0pt}
\begin{minipage}[t][6.3cm][t]{1.0\textwidth} 
  \begin{wrapfigure}{L}{0.25\textwidth}
    \includegraphics[width=0.25\textwidth]{john_smith.eps}
  \end{wrapfigure}
  \noindent
  {\bfseries John Smith} received his BSc (Mathematics) in 2000 from The University of Maryland. His research interests include lasers and optics.
\end{minipage}
\begin{minipage}{1.0\textwidth}
  \begin{wrapfigure}{L}{0.25\textwidth}
    \includegraphics[width=0.25\textwidth]{alice_smith.eps}
  \end{wrapfigure}
  \noindent
  {\bfseries Alice Smith} also received her BSc (Mathematics) in 2000 from The University of Maryland. Her research interests also include lasers and optics.
\end{minipage}
\endgroup
}{}

\end{document}